\documentclass[12pt]{article}

\catcode`@=12 \topmargin -0.5in \evensidemargin 0.0in
\begin{document}
\thispagestyle{empty}
\begin{flushright}
UCRHEP-T392\\
July 2005\
\end{flushright}
\vspace{0.5in}
\begin{center}
{\Large \bf Puzzle of W Leptonic Decay Branching Fractions\\ and
Gauge Model of Generation Nonuniversality \\} \vspace{1.5in} {\bf
Xiao-Yuan Li$^a$\footnote{Email: lixy@itp.ac.cn, Tel:
+(86)-(10)-62551187, Fax: +(86)-(10)-62562587}and Ernest Ma$^b$\\}
\vspace{0.2in} {$^a$ \sl Institute of Theoretical Physics, Chinese
Academy of Sciences, Beijing 100080, China\\} \vspace{0.1in} {$^b$
\sl Physics Department, University of California, Riverside,
California 92521, USA\\}
\vspace{0.8in}
\end{center}
\begin{abstract}
Lepton generation universality holds very well in $Z$ decays, but appears
to be violated in recent LEP data of $W$ leptonic decay branching fractions.
If this trend persists, a consistent and natural explanation is a model
of generation nonuniversality, based on the gauge group $SU(2)_L\times U(1)_R
\times U(1)_{B-L}$.\\
\end{abstract}

Keywords: Gauge extension of the standard model, Lepton generation
nonuniversality

PACS numbers : 12.15.-y 12.60.-i 14.60.-z 14.65.-q

\newpage
\baselineskip 24pt

\section{\bf Introduction}

The principle of universality of the weak interaction is a concept
of deep and enduring significance. The idea arose originally more
than fifty years ago and has been directly and indirectly related
to a number of fundamental developments in particle
physics \cite{sirlin}. The Standard Model of particle interactions
incorporates universality in a fundamental way that all 3
generations of left-handed (right-handed) quarks and leptons are
doublets (singlets) under the same $SU(2)_L \times U(1)_Y$ gauge
group and the couplings are universal parameters. Therefore we may
call the weak interaction universality also the generation
universality.

Test of weak interaction universality in the quark sector is
essentially a test of the CKM unitarity.  Precise measurements
are hitherto confined to the elements of the first row and a definite
conclusion
is not possible before more precise data
become available
\cite{abele2002}\cite{abele2004}\cite{czarnecki2004}.

The universality of the leptonic $Z$ couplings has been accurately
tested at LEP and SLC through a precise analysis of $e^+e^-
\rightarrow \gamma, Z \rightarrow \bar{f} f $ data and it is now
verified to the $0.15\%$ level for the axial vector couplings,
while only a few percent precision has been achieved for vector
couplings \cite{lep1}. For the universality of the leptonic charged
couplings, it can be tested indirectly in $\tau$ decays and the current
data verify it to the $0.2\%$ level \cite{davieryuan}\cite{gan}.

Recently the LEP2 $W-$pair sample has made it possible for the first
direct
measurement of all leptonic $W$ decay branching ratios, i.e.
\cite{lep1}\cite{teubert}\cite{azzurri}\cite{grunewald}
$$ Br(W \rightarrow e \bar{\nu}_e) = 10.65 \pm 0.17\% $$
$$ Br(W \rightarrow {\mu} \bar{\nu}_{\mu}) = 10.59 \pm 0.15\% $$
$$ Br(W \rightarrow {\tau} \bar{\nu}_{\tau}) = 11.44 \pm 0.22\% $$

It should be noted that the branching fractions in taus with
respect to electrons and muons differ by more than two standard
deviations while the branching fractions of $W$ into electrons and
into muons perfectly agree. Assuming only partial lepton
universality the ratio between the tau fractions and the average
of electrons and muons can be computed:
\begin{equation}
\frac{\displaystyle 2 {\cal B}(W \rightarrow \tau
\bar{\nu}_{\tau})}{\displaystyle ({\cal B}(W \rightarrow \mu
\bar{\nu}_{\mu})+{\cal B}(W \rightarrow e \bar{\nu}_{e}))} = 1.077
\pm 0.026
\end{equation}
resulting in a poor agreement at the level of 3.0 standard
deviations, with all correlations included.  If these data
persist, then a puzzle is presented as to whether generation
universality is truly fundamental or not after all. In particular,
one should look for a theoretical framework which allows deviation from
generation universality at a few percent level in the $W$ leptonic
decays while keeping the verified generation universality at 0.2
percent level in the $Z$ leptonic decays.

Perhaps the well tested $e-\mu$ universality and less well tested
$e-\mu-\tau$ universality are only accidental and approximate
symmetries in analogy with flavor $SU(2)$ and $SU(3)$, and owe
their existence to certain mass scale inequalities yet to be
discovered. This theoretically attractive possibility was proposed
by the authors in the so-called gauge model of generation
nonuniversality over twenty years ago \cite{lima81}. With the
discovery of a heavy top quark, the smallness of CKM mixing
angles, and the huge hierarchy in masses, the idea of treating the
third generation differently from the first two generations have
now received much more attention. From the point of view of modern
effective field theory, the conservation of baryon number, lepton
number, the absence of FCNC's and weak universality can be
explained as a consequence of accidental symmetries if the
Standard Model is an effective theory. In fact, in an effective
field theory, the leading term in the expansion often has some new
approximate symmetry \cite{wise}, and the discovery of neutrino
oscillation and neutrino mass can be considered as a first
evidence of this approach. Thus a possible violation of the
universality of weak interactions should not be unexpected.

The phenomenology of the gauge model of generation nonuniversality
was studied in later publications
\cite{malitu88}\cite{mang88}\cite{lima92}\cite{lima93}. The
difference of the branching fractions in taus with respect to
electrons and muons in $W$ leptonic decay, in particular, was
expected in the $ U(1)\times SU(2)_{12} \times SU(2)_3$ model
\cite{lima92}\cite{lima93}, where the subscripts (12) and 3 refer to
1st and 2nd, and 3rd generation of fermions. The same order of the
difference in $Z$ leptonic decay, however, was also expected. In
this paper the lepton universality is reanalyzed in the light of
the recent LEP result of the $W$ leptonic decay branching
fractions in a generation nonuniversality version of the gauge
model $SU(2)_L \times U(1)_R \times U(1)_{B-L}$. It is found that
the observed pattern of lepton universality and nonuniversality in
$Z$ and $W$ decays can be reproduced.

\section{\bf The model}

Our model is based on a generation nonuniversality version of the
gauge model $SU(2)_L \times U(1)_R \times U(1)_{B-L}$, the gauge
group $ SU(3)_c\times SU(2)_{e\mu} \times SU(2)_{\tau} \times
U(1)_{\chi e\mu} \times U(1)_{\chi\tau} \times U(1)_{v e\mu}\times
U(1)_{v\tau}$ with gauge couplings $g_3$, $g_{2e\mu}$,
$g_{2\tau}$, $g_{\chi e \mu}$, $g_{\chi \tau}$,$g_{v e \mu}$,
$g_{v \tau}$, respectively. The fermions are assumed to transform
as follows:
\begin{eqnarray}
&&(u,d)_L \sim (3,2,1,0,0 ,1/3,0), ~~ u_R \sim
(3,1,1,1,0,1/3,0), ~~ d_R \sim (3,1,1,-1,0, 1/3,0); \nonumber \\
&&(\nu,e)_L \sim (1,2,1,0,0,-1,0), ~~ \nu_{eR}  \sim
(1,1,1,1,0,-1,0)~~ e_R \sim (1,1,1,-1,0,-1,0).\nonumber \\
&&(t,b)_L \sim (3,1,2,0,0,0,1/3), ~~ t_R \sim
(3,1,1,0,1,0,1/3), ~~ b_R \sim (3,1,1,0,-1,0,1/3); \nonumber \\
&&(\nu,\tau)_L \sim (1,1,2,0,0,0,-1), ~~ \nu_{\tau R}  \sim
(1,1,1,0,1,0,-1,)~~ \tau_R \sim (1,1,1,0,-1,0,-1).\nonumber \\
\end{eqnarray}

It is understood that the second generation of fermions transforms
as the first generation and all fermions are so-called weak
eigenstates.

The scalar sector consists of four doublets
\begin{eqnarray}
&(\phi_{ee}^+,\phi_{ee}^0) \sim (1,2,1,1,0,0,0), ~~
(\phi_{\tau\tau}^+,\phi_{\tau\tau}^0) \sim
(1,1,2,0,1,0,0),\nonumber \\
&(\phi_{e\tau}^+,\phi_{e\tau}^0) \sim (1,2,1,0,1,-1,1), ~~
(\phi_{\tau e}^+,\phi_{\tau e}^0) \sim
(1,1,2,1,0,1,-1),\nonumber \\
\end{eqnarray}
with the vev (vacuum expectation values) $v_{ee}$, $v_{\tau\tau}$,
$v_{e\tau}$, and $v_{\tau e}$ respectively which break the usual electroweak
symmetry to $U(1)_{em}$ and give masses to the leptons; four singlets
\begin{eqnarray}
& \chi^0 \sim (1,1,1,1,-1,0,0), ~~\upsilon^0 \sim
(1,1,1,0,0,1,-1), \nonumber\\
& \zeta^0_{e\mu} \sim (1,1,1,1,0,-1,0), ~~ \zeta^0_{\tau} \sim
(1,1,1,0,1,0,-1),\nonumber \\
\end{eqnarray}
and one self-dual bi-doublet
\begin{equation}
\eta = {1 \over \sqrt 2} \pmatrix {\eta^0 & -\eta^+ \cr \eta^- &
\bar \eta^0} \sim (1,2,2,0,0,0,0),
\end{equation}
such that $\eta = \tau_2 \eta^* \tau_2$. Each column is a doublet
under $SU(2)_{e\mu}$ and each row is a doublet under
$SU(2)_{\tau}$.

Let $\langle \eta^0 \rangle \equiv u$ (which breaks $SU(2)_{e\mu}
\times SU(2)_{\tau}$ to the usual $SU(2)_L$), $\langle \chi^0
\rangle \equiv z$ (which breaks $U(1)_{\chi e\mu} \times
U(1)_{\chi \tau}$ to $U(1)_{\chi}$,) $\langle \upsilon^0 \rangle
\equiv w$ (which breaks $U(1)_{V e\mu} \times U(1)_{V \tau}$ to
$U(1)_V$), $\langle \zeta^0_{e\mu} \rangle \equiv x$ (which breaks
$U(1)_{\chi e\mu} \times U(1)_{V e\mu}$ to $U(1)_{Y e\mu}$) and
$\langle \zeta^0_{\tau} \rangle \equiv y$ (which breaks
$U(1)_{\chi \tau} \times U(1)_{V \tau}$ to $U(1)_{Y\tau}$).

Instead of the six gauge couplings we will use the electric charge
$e$ and the ratios of coupling constants $\sin^2\theta$,
$\cos^2\phi$, $a$, $b$, and $c$
\begin{eqnarray}
& \frac{\displaystyle 1}{\displaystyle e^2}=\frac{\displaystyle
1}{\displaystyle g^2_0}+\frac{\displaystyle 1}{\displaystyle
G^2}; \nonumber \\
& \frac{\displaystyle 1}{\displaystyle G^2}=\frac{\displaystyle
1}{\displaystyle g^2_{2 e\mu}}+\frac{\displaystyle
1}{\displaystyle g^2_{2\tau}}; ~~\frac{\displaystyle
1}{\displaystyle g_0^2}=\frac{\displaystyle 1}{\displaystyle
g^2_{\chi}}+\frac{\displaystyle 1}{\displaystyle g^2_{V}};
\nonumber \\
& \frac{\displaystyle 1}{\displaystyle
g^2_{\chi}}=\frac{\displaystyle 1}{\displaystyle g^2_{\chi
e\mu}}+\frac{\displaystyle 1}{\displaystyle g^2_{\chi
\tau}};~~\frac{\displaystyle 1}{\displaystyle
g^2_V}=\frac{\displaystyle 1}{\displaystyle
g^2_{Ve\mu}}+\frac{\displaystyle 1}{\displaystyle g^2_{V\tau}};
\nonumber \\
&\frac{\displaystyle e^2}{\displaystyle
G^2}=\sin^2\theta;~~\frac{\displaystyle g^2_0}{\displaystyle
g^2_V}= \cos^2\phi; \nonumber\\
 &\frac{\displaystyle
G^2}{\displaystyle g^2_{2\tau}}=a;~~\frac{\displaystyle
g^2_{\chi}}{\displaystyle g^2_{\chi \tau}}= b;
~~\frac{\displaystyle g^2_V}{\displaystyle
g^2_{V \tau}}= c; \nonumber\\
\end{eqnarray}
Also instead of nine vev's we will use the Fermi constant
$\frac{\displaystyle G_{\mu}}{\displaystyle \sqrt{2}}$ and the
ratios of vacuum expectation values $r$, $s$, $t$,
$1-\frac{\displaystyle 1}{\displaystyle
\xi}$,$1-\frac{\displaystyle 1 }{\displaystyle \kappa }$,$1-
\frac{\displaystyle 1 }{\displaystyle \lambda }$, and
$1-\frac{\displaystyle 1}{\displaystyle \eta}$,
\begin{eqnarray}
&\frac{\displaystyle 4 G_{\mu}}{\displaystyle \sqrt{2} }=
\frac{\displaystyle 1 + (1-\frac{\displaystyle 1}{\displaystyle
\xi})}{\displaystyle (v^2_{\tau e} +
v^2_{\tau\tau})[r(1-\frac{\displaystyle 1}{\displaystyle
\xi})+1+r]} \nonumber \\
&\frac{\displaystyle v^2_{ee}+v^2_{e\tau}}{\displaystyle v^2_{\tau
e} +v^2_{\tau\tau}}= r; ~~\frac{\displaystyle v^2_{ee} +v^2_{\tau
e} }{\displaystyle v^2_{\tau e} +
v^2_{\tau\tau}}=s;~~\frac{\displaystyle v^2_{e\tau} +v^2_{\tau
\tau} }{\displaystyle v^2_{\tau e} + v^2_{\tau\tau}}=t; ~~s+t =
1+r;
\nonumber\\
& 1-\frac{\displaystyle 1}{\displaystyle \xi}=\frac{\displaystyle
v^2_{\tau e}+v^2_{\tau\tau}}{\displaystyle u^2
};~~1-\frac{\displaystyle 1}{\displaystyle
\kappa}=\frac{\displaystyle v^2_{\tau
e}+v^2_{\tau\tau}}{\displaystyle x^2+y^2 };~~1-\frac{\displaystyle
1}{\displaystyle \lambda}=\frac{\displaystyle v^2_{\tau
e}+v^2_{\tau\tau}}{\displaystyle z^2 };~~1-\frac{\displaystyle
1}{\displaystyle \eta}=\frac{\displaystyle v^2_{\tau
e}+v^2_{\tau\tau}}{\displaystyle w^2 };
\nonumber \\
\end{eqnarray}
(Note that $x$ and $y$ appear only in the combination $x^2+y^2$.)
In order to reproduce the observed approximate universality a
hierarchy for vev's

$$ u^2, w^2, z^2, >> x^2+y^2 >> v^2_{ee}, v^2_{e\tau}, v^2_{\tau
e}, v^2_{\tau\tau} $$ is necessary and
$$
\sqrt{v^2_{\tau e} + v^2_{\tau\tau} + v^2_{ee} + v^2_{e \tau}} = 174
~{\rm GeV}
$$
is the electroweak symmetry breaking scale.

\section{\bf The effective charged-current four-fermion weak interaction at low
energy}

The effective charged-current four-fermion weak interaction at low
energy is given by \cite{lima93}
$$
{\cal L}^{cc} = \frac{\displaystyle G_{\mu}}{\displaystyle
\sqrt{2}}\left(
\begin{array}{ll}
J^+_{l}& J^+_{\tau}\\
\end{array}\right)\left( \begin{array}{lc}
                  1& \frac{\displaystyle 1}{\displaystyle \xi}\\
                  \frac{\displaystyle 1}{\displaystyle \xi}&1 +(
                  1-\frac{\displaystyle 1}{\displaystyle
                  \xi})(r-1)\\
                  \end{array}\right)\left(\begin{array}{ll}
                                          J^-_{l}\\
                                          J^-_{\tau}\\
                                          \end{array}\right)
$$
where
$$ J^-_{l} = \bar{\nu}_l \gamma^{\alpha}(1-\gamma_5)l
+ \bar{\nu}_{\mu}\gamma^{\alpha}(1-\gamma_5)\mu
$$
$$
J^-_{\tau} = \bar{\nu}_{\tau} \gamma^{\alpha}(1-\gamma_5)\tau
$$

Thus the $\tau$ leptonic decay width is given by
\begin{equation}
\Gamma(\tau \rightarrow l \bar{{\nu}_l}\nu_{\tau})= (\Gamma(\tau
\rightarrow l \bar{{\nu}_l}\nu_{\tau}))^{\rm SM}( 1
-2(1-\frac{\displaystyle 1}{\displaystyle \xi})),
\label{taudecay}
\end{equation}
where $l=e, \mu$ and $(\Gamma(\tau \rightarrow l
\bar{{\nu}_l}\nu_{\tau}))^{\rm SM}$ can be calculated from
experimental measurements of $\mu$ decay rate, $m_{\mu}$,
$m_{\tau}$, $M_W $ and
$\alpha^{-1}(m_{\tau})$\cite{MarcianoSirlin},
$$
(\Gamma(\tau \rightarrow l \bar{{\nu}_l}\nu_{\tau}))^{\rm SM}=
\frac{\displaystyle G^2_{\mu} m^5_{\tau}}{\displaystyle
192\pi^3}f(\frac{\displaystyle m^2_l}{\displaystyle m^2_{\tau}})
(1 + \frac{\displaystyle 3m^2_{\tau} }{\displaystyle 5 M^2_W
})(1+\frac{\displaystyle \alpha(m_{\tau}) }{\displaystyle 2\pi
}(\frac{\displaystyle 25}{\displaystyle 4}-\pi^2))
$$
where
$$
f(x)= 1 - 8x + 8x^3 - x^4 -12x\ln x
$$
The $e -\mu -\tau$universality can be tested by comparing the
above theoretical prediction with the measurements of $\tau$
lifetime and lepton decay branching ratios
$$\frac{\displaystyle Br(\tau \rightarrow l \bar{\nu_l}
\nu_{\tau})}{\displaystyle t_{\tau}}$$

\section{\bf The lightest W and Z bosons}

Due to the mixing with the extra gauge bosons the masses of the
lightest $W$ and $Z$ get the shift
$$
M^2_W = \frac{\displaystyle \pi \alpha}{\displaystyle
\sqrt{2}G_{\mu}}\frac{\displaystyle 1}{\displaystyle \sin^2\theta
(1 + \Delta\rho_C) }
$$
\begin{equation}
\Delta\rho_C = -(1-\frac{\displaystyle 1}{\displaystyle
\xi})a[2-(1+r)a]
\end{equation}
$$
M^2_Z= \frac{\displaystyle \pi \alpha}{\displaystyle \sqrt{2}
G_{\mu} }\frac{\displaystyle 1}{\displaystyle \sin^2\theta
\cos^2\theta(1 + \Delta\rho_N)}
$$
\begin{equation}
\begin{array}{lcl}
\Delta\rho_N &=& \Delta\rho_C\\
              && +(1-\frac{\displaystyle 1}{\displaystyle
              \kappa})\cos^4\phi(1+r)\\
              && +(1-\frac{\displaystyle 1}{\displaystyle \lambda})\frac{\displaystyle [t-b(s+t)]^2}{\displaystyle
              1+r}\\
              && +(1-\frac{\displaystyle 1}{\displaystyle \eta})\frac{\displaystyle (r-s)^2}{\displaystyle
              1+r}\\
\end{array}
\end{equation}
By identifying the lightest $Z$ as the observed $Z$ the $\sin^2\theta$
can be related to the $(\sin^2\theta_W)^{\rm SM}$ \cite{bgklm}
\begin{equation}
\sin^2\theta = (\sin^2\theta_W)^{\rm SM}( 1- \frac{\displaystyle
1-s^2_0 }{\displaystyle 1-2 s^2_0}\Delta\rho_N)
\end{equation}
where $s_0^2$ is defined as usual by
$$
s_0^2 (1-s_0^2) = {\pi \alpha (M_Z) \over \sqrt 2 G_{\mu} M_Z^2}.
$$

Thus the mass of the observed W is
\begin{equation}
M^2_W = (M^2_W)^{\rm SM}(1-\Delta\rho_C  + \frac{\displaystyle
1-s^2_0 }{\displaystyle 1-2s^2_0}\Delta\rho_N)
\label{Wmass}
\end{equation}
and its charged-current interaction is given by
$$
{\cal L}^{cc}_W = \frac{\displaystyle 1}{\displaystyle
2\sqrt{2}}\frac{\displaystyle e}{\displaystyle \sin\theta
}[(1+\Delta g^{l}_{\xi}) J^-_{l} + (1 + \Delta g^ {\tau}_{\xi})
J^-_{\tau}]W^+  + {\rm h.c.}
$$
with
\begin{equation}
\Delta g^{l}_{\xi} = (1 - \frac{\displaystyle 1}{\displaystyle
\xi}) a [ 1-(1+r)a]
\end{equation}
\begin{equation}
\Delta g^{\tau}_{\xi} = -(1 - \frac{\displaystyle 1}{\displaystyle
\xi})(1- a) [ 1-(1+r)a]
\end{equation}
Therefore the couplings of the lightest $W$ to $l\bar{\nu}_l$ and
$\tau\bar{\nu}_{\tau}$ are
$$
g^{W l \bar{\nu}_l} = (g^{W l \bar{\nu}_l})^{\rm SM} ( 1 +
\frac{\displaystyle 1}{\displaystyle 2} \frac{\displaystyle
1-s^2_0}{\displaystyle 1-2s^2_0} \Delta \rho_N + \Delta g^l_{\xi})
$$
$$
g^{W \tau \bar{\nu}_{\tau}} = (g^{W \tau \bar{\nu}_{\tau}})^{\rm
SM} ( 1 + \frac{\displaystyle 1}{\displaystyle 2}
\frac{\displaystyle 1-s^2_0}{\displaystyle 1-2s^2_0} \Delta \rho_N
+ \Delta g^{\tau}_{\xi})
$$
respectively and the coupling ratio is
\begin{equation}
\frac{\displaystyle g^{W \tau \bar{\nu}_{\tau}}}{\displaystyle
g^{W l \bar{\nu}_l}} = (\frac{\displaystyle g^{W \tau
\bar{\nu}_{\tau}}}{\displaystyle g^{W l \bar{\nu}_l}})^{\rm
SM}(1+\Delta g^{\tau}_{\xi}- \Delta g^{l}_{\xi})
\label{Wdecay}
\end{equation}
To the leading order, the finite fermion mass effects in the $W$
leptonic partial decay widths can be neglected and we obtain
$$ \Gamma_{W
l\bar{\nu}} = (\Gamma_{W l \bar{\nu}})^{\rm SM} ( 1 +
\frac{\displaystyle 3}{\displaystyle 2}\frac{\displaystyle c^2_0}
{\displaystyle c^2_0 -s^2_0}\Delta \rho_N
- \frac{\displaystyle 1}{\displaystyle 2}\Delta \rho_C + 2\Delta g^l_{\xi}) \\
\label{Wwidth}
$$

Using $s^2_0 = 0.23$ we have
\begin{equation}
\Gamma_{W l\bar{\nu}_l} = (\Gamma_{W l \bar{\nu}_l})^{\rm SM} ( 1
+ 2.139 \Delta \rho_N
- 0.500 \Delta \rho_C + 2\Delta g^{l}_{\xi}) \\
\label{wenu}
\end{equation}
\begin{equation}
\Gamma_{W \tau \bar{\nu}_{\tau}} = (\Gamma_{W \tau
\bar{\nu}_{\tau}})^{\rm SM} ( 1 + 2.139 \Delta \rho_N
- 0.500 \Delta \rho_C + 2\Delta g^{\tau}_{\xi}) \\
\label{wtaunu}
\end{equation}

The neutral-current interaction of the lightest $Z$ can be written
as
\begin{equation}
{\cal L}^{\rm NC} = (\sqrt{2} G_{\mu} M^2_Z)^{1/2}
\bar{f}\gamma^{\alpha}(g_{vf}-g_{af}\gamma_5)f Z_{\alpha}
\end{equation}
with the effective axial-vector and vector couplings
\begin{equation}
g_{af}=(\rho^f)^{1/2} I_3(f)
\end{equation}
\begin{equation}
g_{vf}=(\rho^f)^{1/2} (I_3(f)- 2Q(f) \sin^2\theta^f_{\rm eff})
\end{equation}
where
$$
\rho^f =(\rho^f)^{\rm SM} +\Delta\rho^f
$$
$$
\sin^2\theta^f_{\rm eff} =(\sin^2\theta^f_{\rm eff})^{\rm SM}
+\Delta\sin^2\theta^f_{\rm eff}
$$
For the charged leptons $l$ we have
$$
\Delta \rho^l = \Delta\rho_N + 2\Delta g^l_{\xi} + 2\Delta
g^l_{\alpha}
$$
$$
\Delta\sin^2\theta^l_{\rm eff} = -\frac{\displaystyle c^2_0 s^2_0
}{\displaystyle c^2_0 -s^2_0}\Delta \rho_N -s^2_0 \Delta g^l_{\xi}
+ c^2_0 \Delta g^{l}_{\alpha} -\Delta
g^l_{\beta}\frac{\displaystyle V(l)}{\displaystyle 2Q(l)}
$$
for the charged lepton $\tau$ we have the similar equations with
replacement $l \rightarrow \tau$.

For neutrinos $\nu_l$
$$
\begin{array}{lcl}
\Delta \rho^{\nu_l} &=& \Delta\rho_N + 2\Delta g^l_{\xi} + 2\Delta
g^{l}_{\alpha} + 2\Delta g^{l}_{\beta}
\frac{\displaystyle V(\nu_l)}{\displaystyle 2I_3(\nu_l)}\\
\end{array}
$$
where we have assumed the seesaw mechanism for neutrino masses.
Also the replacement $l \rightarrow \tau$ is used for
$\nu_{\tau}$.

Notice that $\Delta\rho_N$ comes from the mass shift of the
lightest $Z$ due to mixing with extra $Z$ bosons,
$\frac{\displaystyle c^2_0}{\displaystyle
c^2_0-s^2_0}\Delta\rho_N$ is the shift from $\sin^2\theta$,
$\Delta g$'s are the coupling shifts of the lightest $Z$ due to
mixing with extra $Z$ bosons.
\begin{equation}
\Delta g^{l}_{\alpha} = -(1 - \frac{\displaystyle 1}{\displaystyle
\kappa})(1+r)\cos^4\phi -(1 - \frac{\displaystyle 1}{\displaystyle
\lambda}) b [ t-(s+t)b]
\end{equation}
\begin{equation}
\Delta g^{\tau}_{\alpha} = -(1 - \frac{\displaystyle
1}{\displaystyle \kappa})(1+r)\cos^4\phi +(1 - \frac{\displaystyle
1}{\displaystyle \lambda}) (1-b) [ t-(s+t)b]
\end{equation}
\begin{equation}
\begin{array}{lcl}
\Delta g^{l}_{\beta}&=& -(1 - \frac{\displaystyle
1}{\displaystyle \kappa})(1+r)\cos^2\phi  \\
&& -(1 - \frac{\displaystyle 1}{\displaystyle \lambda}) b [
t-(s+t)b]
 -(1 - \frac{\displaystyle 1}{\displaystyle \eta}) c (r-s)\\
\end{array}
\end{equation}
\begin{equation}
\begin{array}{lcl}
\Delta g^{\tau}_{\beta}&=& -(1 - \frac{\displaystyle
1}{\displaystyle \kappa})(1+r)\cos^2\phi  \\
&& +(1 - \frac{\displaystyle 1}{\displaystyle \lambda}) (1-b) [
t-(s+t)b]+(1 - \frac{\displaystyle 1}{\displaystyle \eta})(1- c)
(r-s)\\
\end{array}
\end{equation}

From $\Delta\rho^f$, $\Delta\sin^2\theta^f_{eff}$ and the
effective vector and axial-vector couplings of the Standard Model
we can obtain the $Zf\bar{f}$ partial decay width
$$
\Gamma^f = (\Gamma^f)^{\rm SM} ( 1 + \Delta \rho^f
-\frac{\displaystyle 4g_{vf} Q(f) } {\displaystyle g^2_{vf} +
g^2_{af}}\Delta \sin^2\theta^f_{eff}),
$$
the asymmetry
$$
{\cal A}^f = ({\cal A}^f)^{\rm SM} + \frac{\displaystyle 4g_{af}
Q(f) [g^2_{vf} - g^2_{af}]} {\displaystyle [g^2_{vf} +
g^2_{af}]^2}\Delta\sin^2\theta^f_{eff},
$$
and the forward backward asymmetries
$$
{\cal A}^{0,f}_{\rm FB}= ({\cal A}^{0,f}_{\rm FB})^{\rm SM} +
\frac{\displaystyle 3 }{\displaystyle 4 }[ ({\cal A}_e)^{\rm SM}
\Delta {\cal A}_f + ({\cal A}_f)^{\rm SM} \Delta {\cal A}_e]
$$
where $ g_{vf}$, $g_{af}$ and $Q(f)$ are vector coupling, axial
vector coupling and electric charge of fermion $f$ in the SM
respectively.

It should be emphasized that the description in terms of the 
gauge-boson mass shifts $\Delta\rho$'s and coupling shifts $\Delta g$'s
is quite general in the sense that it depends only on the gauge
group structure and the assignments of fermion quantum numbers. In
this context, a few remarks are in
order:

\begin{itemize}
\item  Because of the weak isospin symmetry the large
nonuniversality effects in $W$ charged lepton decays will induce 
sizable nonuniversality effects in $Z$ decays as well. In order to
maintain the verified generation universality at 0.2 percent level
in $Z$ leptonic decays, two isospin singlet nonuniversally coupled
neutral gauge bosons must be introduced with chiral and vector
couplings respectively. This leads to our generation
nonuniversality version of the gauge model $SU(2)_L \times U(1)_R
\times U(1)_{B-L}$.
\item  In this model however,  $\Delta\sin^2\theta^f_{\rm eff}$,
the shift of $\sin^2\theta^f_{\rm eff}$, does depend on the
electric charge $Q(f)$ and $B-L$ charge $V(f)$ of a given fermion
$f$ (charged leptons, up and down quarks) although $\Delta\rho^f$,
the shift of $\rho^f$, does not. Therefore the nonuniversality
effects at the $Z$ pole for charged leptons, up and down quarks
cannot be minimized simultaneously.  Sizable nonuniversality
effects in the quark sector are thus expected.

\item In contrast to the SM, quark mixing in our gauge model
of generation nonuniversality has more unknown parameters
because the Cabibbo-Kobayashi-Maskawa matrix is no longer
unitary.  At this time it is not possible to use the experimental
data of the quark sector to put meaningful constraints on the model.

\item The shift $\Delta\rho^{\nu_l}$ of $\rho^{\nu_l}$,
also depends on the third component of weak isospin $I_3(\nu_l)$ and the
$B-L$ charge $V(\nu_l)$ of neutrinos. As a result, significant
nonuniversality effects in the neutrino sector may also be expected in
this model.

\item 
However, as shown below, the nonuniversality
effects at the $Z$ pole for charged leptons and neutrinos can be
minimized simultaneously.  
This allows us to explain the
experimental result that the number of light neutrino species
given by the ratio of the $Z$ decay width into invisible particles
$\Gamma_{inv}$ and the leptonic decay width $\Gamma_{l\bar{l}}$ is only
slightly below (by two standard deviations) the value of 3 expected from 3
observed fermion families \cite{lep1}.
\end{itemize}

Taking $s^2_0 = 0.23$, in particular, we have
\begin{equation}
\Gamma_{inv} = (\Gamma_{inv})^{SM} [1 + \Delta \rho_N
+\frac{\displaystyle 4}{\displaystyle 3}(\Delta g^{l}_{\xi} +
\Delta g^{l}_{\alpha} -\Delta g^{l}_{\beta}) +\frac{\displaystyle
2}{\displaystyle 3}(\Delta g^{\tau}_{\xi} + \Delta
g^{\tau}_{\alpha}- \Delta g^{\tau}_{\beta})]
\end{equation}
\begin{equation}
\begin{array}{lcll}
\Gamma_{l \bar{l}} &=& (\Gamma_{l \bar{l}})^{\rm SM}&(1 + 1.209
\Delta\rho_N + 2.146 \Delta g^{l}_{\xi} + 1.510\Delta
g^{l}_{\alpha}+0.318\Delta g^{l}_{\beta})\\
\end{array}
\end{equation}
\begin{equation}
\begin{array}{lcll}
\Gamma_{\tau \bar{\tau}} &=& (\Gamma_{\tau \bar{\tau}})^{\rm
SM}&(1 + 1.209 \Delta\rho_N + 2.146 \Delta g^{\tau}_{\xi} +
1.510\Delta
g^{\tau}_{\alpha}+0.318\Delta g^{\tau}_{\beta})\\
\end{array}
\end{equation}
\begin{equation}
\begin{array}{lcll}
{\cal A}_l &= & ({\cal A}_l)^{\rm SM} &+ 2.574 \Delta\rho_N +1.805
\Delta g^{l}_{\xi} - 6.043\Delta g^{l}_{\alpha}+ 3.924\Delta
g^{l}_{\beta}\\
\end{array}
\end{equation}
\begin{equation}
\begin{array}{lcll}
{\cal A}_{\tau} &= & ({\cal A}_{\tau})^{\rm SM} &+ 2.574
\Delta\rho_N + 1.805 \Delta g^{\tau}_{\xi} - 6.043\Delta
g^{\tau}_{\alpha}+ 3.924\Delta
g^{\tau}_{\beta}\\
\end{array}
\end{equation}
\begin{equation}
\begin{array}{lcll}
{\cal A}^{0,l}_{\rm FB} &=& ({\cal A}^{0,e}_{\rm FB})^{\rm SM}& +
0.614 \Delta\rho_N  + 0.430 \Delta g^{l}_{\xi}- 1.442\Delta
g^{l}_{\alpha}
+0.936\Delta g^{l}_{\beta}\\
\end{array}
\end{equation}
\begin{equation}
\begin{array}{lcll}
{\cal A}^{0,\tau}_{\rm FB} &=&({\cal A}^{0,\tau}_{\rm FB})^{\rm
SM}& + 0.614 \Delta\rho_N + 0.215 \Delta g^{l}_{\xi} + 0.215 \Delta g^{\tau}_{\xi}\\
&&&- 0.721\Delta g^{l}_{\alpha} - 0.721\Delta g^{\tau}_{\alpha}
+0.468 \Delta g^{l}_{\beta} +0.468 \Delta g^{\tau}_{\beta}\\
\end{array}
\end{equation}
\begin{table}[htb]
\begin{center}
\begin{tabular}{|l|l|l|l|l|l|}
\hline
             & \parbox{2.5cm}{Measurement with Total Error} &\parbox{1.0cm}{Standard Model fit}& Pull& This model & Pull\\
       \hline
       \parbox{5.0cm}{LEP} &&&&&\\
       \parbox{5.0cm}{{\sf line-shape and lepton asymmetries}} &&&&&\\
       $\Gamma_{ee}({\rm MeV})$ & $83.92 \pm 0.12 $ & 84.036& 0.97& 83.94 & 0.19\\
       $\Gamma_{\mu \mu}({\rm MeV})$ & $83.99 \pm 0.18 $ & 84.036& 0.26& 83.94 &-0.26\\
       $\Gamma_{\tau\tau}({\rm MeV})$ & $84.08 \pm 0.22 $ & 84.036 & -0.20 &84.08 & 0.01\\
       $\Gamma_{\rm inv}({\rm GeV})$ & $ 0.4974 \pm 0.0025 $ & 0.5017 & 1.72& 0.4978 & 0.15\\
       $ A^{0,e}_{\rm FB}$ & $0.0145 \pm 0.0025$ & 0.0165 & 0.80& 0.0174 & 1.14\\
       $ A^{0,\mu}_{\rm FB}$ & $0.0169 \pm 0.0013$ & 0.0165 & -0.31& 0.0174 & 0.35\\
       $ A^{0,\tau}_{\rm FB}$ & $ 0.0188 \pm 0.0017$ & 0.0165& -1.35& 0.0165 & -1.34\\
       \parbox{5.0cm}{{\sf $\tau$ polarization}} &&&&&\\
       ${\cal A}_{e}$ & $ 0.1498 \pm 0049 $ & 0.1483 & -0.31& 0.1518 & 0.41\\
       ${\cal A}_{\tau}$ & $ 0.1439 \pm 0043 $ & 0.1483 & 1.02&0.1450 &0.25\\
       \hline
       \parbox{5.0cm}{{\sf polarized lepton asymmetry at SLC}} &&&&&\\
       ${\cal A}_{LR} $ & $ 0.1524 \pm 0.0022 $ & 0.1483 & -1.86& 0.1518 & -0.27\\
       ${\cal A}_e $ & $ 0.1544 \pm 0.0060 $ & 0.1483 & -1.02& 0.1518 & -0.43\\
       ${\cal A}_{\mu} $ & $ 0.142 \pm 0.015 $ & 0.1483 & 0.42& 0.1518 & 0.65\\
       ${\cal A}_{\tau} $& $ 0.136 \pm 0.015 $ & 0.1483 &0.82& 0.1450 & 0.60\\
       \hline
       \parbox{5.0cm}{{\sf W mass}} &&&&&\\
       $ M_W ({\rm GeV})$  & $ 80.4250 \pm 0.0340 $ & 80.3940& -0.91& 80.4258 & 0.03\\
       \parbox{5.0cm}{{\sf W leptonic partial width}} &&&&&\\
       $\Gamma(W \rightarrow e\bar{\nu}_e)({\rm GeV})$ & $ 0.2272 \pm
       0.0082 $ & $0.2267$ & -0.06 & 0.2207 & -0.78\\
       $\Gamma(W \rightarrow \mu\bar{\nu}_{\mu})({\rm GeV})$ & $ 0.2259 \pm
       0.0080 $ & $0.2267$ & 0.10 & 0.2207 & 0.64\\
       $\Gamma(W \rightarrow \tau \bar{\nu}_{\tau})({\rm GeV})$ &
        $0.2440 \pm 0.0092$ & $0.2267$ & -1.88 &  0.2363 & -0.83 \\
       \hline
\end{tabular}
\caption{{\bf  Fit Values of 17 High $Q^2$ Leptonic Observables.}
The measurements with total errors are quoted from references
\cite{lep1}and \cite{grunewald}. The SM values of the
$\Gamma_{ll}$ and $\Gamma_{\rm inv}$ are calculated from the SM
values of $M^2_Z$, $\Gamma_Z$ and $R^0_l$. $W$ leptonic partial
decay widths are calculated from $Br(W \rightarrow l\nu_l)$ and
$\Gamma_W({\rm GeV})$.}
\end{center}
\end{table}

\section{\bf Constraints from present data}

At present the high $Q^2$ data available for testing
lepton universality are the mass and leptonic branching ratios of
the charged $W$ boson
\cite{lep1}\cite{teubert}\cite{azzurri}\cite{grunewald}, the
neutral $Z$ leptonic decay widths and asymmetries\cite{lep1}. A
global fit (using MUNUIT program) to these measurements is
performed and the results, including the LEP and SLD measurements
used, the fit values and pull of the Standard Model, and the fit
values and pull of this model are summarized in Table 1. The
$\chi^2/d.o.f$ of our fit is 6.38/9 to be compared with 17.44/17
in the Standard Model \cite{lep1} which, however, does not
maximize the discrepancy in the $W$ leptonic branching fractions
of Eq.~(1). If subsequent maximization is made (instead, fit to
the ratios $Br(W\rightarrow \mu \bar{\nu}_{\mu})$/$
Br(W\rightarrow e \bar{\nu}_{e})$, $Br(W\rightarrow \tau
\bar{\nu}_{\tau})$/$ Br(W\rightarrow e \bar{\nu}_{e})$ and
$Br(W\rightarrow \tau \bar{\nu}_{\tau})$/$ Br(W\rightarrow \mu
\bar{\nu}_{\mu})$), then ours is 4.74/9 to be compared with
28.66/17 in the Standard Model.

The best-fit values of our parameters are functions of $
\Delta\rho_N$ and are therefore not unique.  One example is given
below:

\begin{equation}
\begin{array}{ll}
\Delta \rho_C = -0.0008 \pm 0.0008;
& \Delta \rho_N = 0.0000;\\
\Delta g^{l}_{\xi}= -0.0133 \pm  0.0079 ; &
\Delta g^{\tau}_{\xi}-\Delta g^{l}_{\xi} = 0.0344\pm  0.0220 ;\\
\Delta g^{l}_{\alpha}=  0.0127  \pm 0.0079 ; &
\Delta g^{\tau}_{\alpha}-\Delta g^{l}_{\alpha} =  -0.0333 \pm 0.0216; \\
\Delta g^{l}_{\beta}= 0.0265 \pm 0.0145;
& \Delta g^{\tau}_{\beta}-\Delta g^{l}_{\beta}= -0.0687 \pm 0.0411.\\
\end{array}
\end{equation}

The interesting point is that the central values of the best fit
parameters satisfy
$$
\Delta g^{\tau}_{\alpha}-\Delta g^{l}_{\alpha} \sim
\frac{1}{2}(\Delta g^{\tau}_{\beta}-\Delta g^{l}_{\beta})\sim
-(\Delta g^{\tau}_{\xi}-\Delta g^{l}_{\xi})\sim {3}\Delta
g^{l}_{\xi}\sim -{3}\Delta g^{l}_{\alpha}\sim
-\frac{3}{2}\Delta g^{l}_{\beta}
$$
which are determined by the group structure and fermion
assignments of the model.

 The above constraints when combined with
low-$Q^2$ measurements such as the low energy $\tau$ decays
\cite{davieryuan}\cite{gan}, the low energy $\nu_{\mu} e$
scattering \cite{PDG} and the polarized M\o ller scattering
\cite{slac158}, should then determine more of our model
parameters and make predictions of the properties of the extra
charged and neutral gauge bosons and other rich phenomenology like
$\bar{\nu}_e e $ scattering \cite{conrad}\cite{rosner},
$e^+e^-\rightarrow\mu^+\mu^-$, $e^+e^-\rightarrow \tau^+\tau^-$
etc. We shall leave that for future publications \cite{lima05}.

\section{\bf Conclusion}

The recent LEP data of the $W$ leptonic decay branching fractions
provide the first evidence for possible violation of
lepton generation universality. It is shown that this deviation from
lepton universality in $W$ leptonic decays and the observed
approximate lepton universality in $Z$ decays result
consistently and naturally in a generation nonuniversality
version of the gauge model $SU(2)_L\times U(1)_R \times
U(1))_{B-L}$.


The work of X.L. was supported in part by the China National
Natural Science Foundation under Grants No.~90103017 and 10475106.
The work of E.M. was supported in part by the U.~S.~Department of
Energy under Grant No.~DE-FG03-94ER40837.

\bibliographystyle{unsrt}

\end{document}